# 3D super-resolved multi-angle TIRF via polarization modulation

CHENG ZHENG,[1,4] GUANGYUAN ZHAO,[1] WENJIE LIU,[1] YOUHUA CHEN,[1] ZHIMIN ZHANG,[1] LUHONG JIN,[2] YINGKE XU,[2] CUIFANG KUANG,[1,3*] AND XU LIU,[1,3]

[1]State Key Laboratory of Modern Optical Instrumentation, College of Optical Science and Engineering, Zhejiang University, Hangzhou 310027, China
[2]Key Laboratory for Biomedical Engineering of Ministry of Education, College of Biomedical Engineering, Zhejiang University, Hangzhou 310027,China
[3]Collaborative Innovation Center of Extreme Optics, Shanxi University, Taiyuan 030006, China
[4]chengzh0427@gmail.com
*Corresponding authors: cfkuang@zju.edu.cn

**Measuring the three dimension nanoscale organization of protein or cellular structures is challenging, especially when the structure is dynamic. Owing to the informative total internal reflection fluorescence (TIRF) imaging under varied illumination angles, multi-angle (MA) TIRF has been examined to offer a nanoscale axial and a sub-second temporal resolution. However, conventional MA-TIRF still performs badly in lateral resolution and fail to characterize the depth-image in densely-distributed regions, leaving a huge contrast between the nanoscale axial resolution and the diffraction limited lateral resolution. Moreover, the previous reconstructions are highly restricted by the efficiency with the increased amount of illumination angles. Here, we for the first time, emphasize the lateral super-resolution in the MA-TIRF and exampled by simply introducing polarization modulation into the illumination procedure. Equipped with a sparsity and accelerated proximal algorithm, we examine a more precise 3D sample structure compared with previous methods, enabling live cell imaging with temporal resolution of 2 seconds, recovering high-resolution mitochondria fission and fusion process. Since the introduced vortex half wave retarder is an add-on component to the existing MA-TRIF system and the algorithm is the first open sourced and the fastest, we anticipate that the method and algorithm introduced here would be adopted rapidly by the biological community.**

## 1. INTRODUCTION

Three dimensional imaging is of great importance in observing the biostructures and their corresponding activities. However, the wave nature of light restricts the 3D resolution, making the details of subcellular structures and protein assemblies unresolvable [1]. To enhance axial resolution/sectioning ability, the initial promotions include the confocal [2] and deconvolved wide-field microscopes [3], promoting viable and fast ways for common implementations. To further tackle the weakness of resolution in these two basic foundations, various super-resolved adaptions have been proposed and examined, such as the multi-photon microscope [4-6], the 3D localization techniques (DNA-PAINT [7, 8], STROM [9, 10]), the 3D stimulated depletion emission microscopes (STED) [11]/ reversible saturable (or switchable) optical linear (fluorescence) transitions (RESOLFT) microscopes [12, 13] as well as the 3D structured illumination microscopes (3D-SIM) [14, 15]. Of all the above explorations, localization microscopes achieves a state-of-the-art resolution up to 5 to 10 nanometers in lateral dimension using the DNA-PAINT technique [16] and 10 to 20 nanometers in axial dimension using the 4Pi scheme [17]. Though promoting a high resolution, the temporal resolution is relative low (from 10 minutes to hours). Many problems emerged due to the large acquisition time, including sample drift, photon bleaching, and more vitally, sample's motion artifact, especially in live cells. Upon this aspect, 3D-SIM is a good alternative in 3D imaging, achieving a time resolution of 5-25s as well as a spatial resolution of 120-nm lateral and 360-nm axial [15]. However, the system burden is relative high because of the interference system while the axial resolution is still not satisfactory.

Being the fastest way of providing sub-diffractive axial imaging of biological samples, total internal reflection fluorescence (TIRF) microscopy is a powerful tool in studying cellular structures at molecule level [18-22]. By further mathematically mapping the varied illumination angles to the penetration depth of evanescent excitation wave, quantitative axial profile is obtained through computational algorithms, dubbed as multi-angle illumination TIRF (MA-TIRF) [23-30]. In the past, we have witnessed various advances in the reconstruction algorithms, including line fitting [23, 25], a four-layer model based linear regression [24], and nonlinear least-squares fitting [26] but with varied restrictions. While in 2014, Boulanger *et al.* overcomes the previous drawbacks by treating MA-TIRF as an ill-posed inverse problem and applied several regularizations with optimization algorithm to realize axial super resolution [27] rendering a compatibility with complex 3D volume samples as well as a high temporal of 7fps and an axial resolution up to 50 nm.

Owing to the above advances, MA-TRIF proves to be more suitable for fast imaging scenarios, such as dynamic events near the plasma membrane. However, though numerous algorithms have been proved, few explorations have paid attention to the system's enhancement, such as addressing the lateral super-resolution, partially because of the huge expense and difficulty of introducing a compatible method into the current MA-TIRF system while still maintaining a high temporal resolution. However, the wide-field scale lateral resolution of conventional TIRF microscopy is not sustainable for discerning the densely labeled regions, leaving such clusters a challenge to be further characterized, i.e., the out-of-proportion contrast between a lateral resolution of 300 nm with an axial resolution of 50 nm requires a promotion in lateral resolution. In addition, despite the variants of algorithms, the reconstruction speed hasn't been promoted. This is crucial to MA-TIRF, where such big data generated from tens of angles requires an adapted and efficient algorithm to ensure the computational tractability for possible real-time observation. Finally, a viable open-source algorithm for the MA-TIRF reconstruction has long been desired but not yet available.

Here we propose a new method, termed Polarization TIRF (Pol-TIRF), to overcome the challenges mentioned above. Inspired by the previous examination of polarization-rendering sparsity[31] for achieving super-resolution in both wide-field [32, 33] and point scanning microscopes [34], we simply introduce a polarization modulator (i.e., a vortex half-wave retarder) into the illumination path of a conventional scanning MA-TIRF system. When altering the polarization of the incident light, the emission intensity of the dipoles are modulated by their relative angle to the incident polarization [35], which adds sparsity to the recordings. Besides promoting this informative acquisition, we substantially enhance the reconstruction accuracy and speed by equipping a computationally efficient algorithm in demodulating the image stack to render an isotropically super-resolved reconstruction.

This paper goes as follows. Firstly, we formulate the principle of Pol-TIRF, mathematically giving the image formation model as well as the reconstruction algorithm. Secondly, we demonstrate that our algorithms outperforms the contemporary ones [27, 30] by simulating the reconstruction result of fluorescent beads and microtubules under various illumination angle numbers and the raw data noise levels. Thirdly, experimental results on fixed cells of microtubules and nuclear pore complexes are evaluated to further prove the Pol-TIRF's superiority in characterizing the lateral features as well as with an enhanced axial resolution. Finally, we show a specific application by successfully capturing the active dynamics of mitochondria in a live U2OS cell. To the best of our knowledge, it is the first time that the consecutive dynamic processes of mitochondrial fission and fusion in live cells are observed by 3D high-resolution imaging while keeping a time resolution of 2s.

## 2. METHOD

### A. Experimental setup

Our Pol-TIRF system is built on a Nikon Eclipse Ti microscope. The control of illumination beam is realized with a self-built two-axis scanning galvanometer [Fig. 1(a)], whose scanning trajectory is programmed with a self-written software. By changing the position that beam illuminates on the vortex half-wave retarder (WPV10L-633, Thorlabs) using the scanning galvanometer, we achieved the modulating of excitation beam polarization state. Shown in Fig. 1(b), this retarder has a constant retardance of $\lambda/2$ across the whole aperture while its fast axis rotates continuously over the area of the optic, with a relationship given as:

$$\theta = \varphi/2 + \delta \quad (1)$$

where $\theta$ is the orientation of the fast-axis of a given azimuthal angle $\varphi$ on the retarder, $\delta$ is the orientation of the fast axis at $\varphi = 0$. When the p-polarized incidence falls on the retarder after the scanning galvanometer, the output polarization is rotated with an angle of $2\gamma$, where $\gamma$ is the relative angle between input polarization and the fast axis of the retarder at the illumination spot. The output polarization state at different azimuthal position when the input polarization is parallel to the fast axis at $\varphi = 0$, is depicted in Fig. 1(b) step 1. Relayed by a 4-f system consisting of L2 and TL, the polarization beam is refocused onto the back focal plane of the microscope objective (Apo 100x 1.49 oil immersion TIRF objective, Nikon). Finally, the oblique illumination excites the fluorescent sample while the emission fluorescence from the sample passes through the dichroic mirror (ZT647 rdc-uf3, Chroma) and gets detected by an Electron-Multiplying CCD (EMCCD, iXon Ultra 888, Andor).

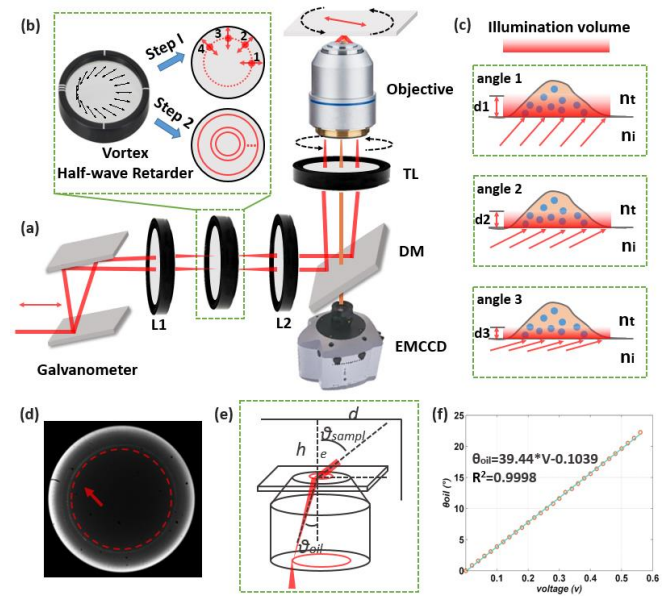

Fig. 1. Experimental setup of Pol-TIRF. (a) Polarization modulation is realized with a vortex half-wave retarder in the dotted green box. L1, L2: Lens; DM: Dichroic Mirror; TL: Tube Lens. (b) The front view of the vortex half-wave retarder, where the black arrows indicate its fast axis orientation distribution. The illumination spot position variation during the two-step acquisition process is also pictured. The red arrows on the spots in step 1 indicates the polarization state after the retarder when input polarization is parallel to the fast axis at $\varphi = 0$; (c) TIRF illumination volume under different illumination angles. $n_t$ and $n_i$ are refractive index of incident and transmission medium; (d) Circles described by the focal spot on the back focal plane; (e) Sketch showing how to calibrating the voltage on the scanning galvanometer and corresponding illumination angle onto the sample; (f) Curve fitting results of the linear relationship between the voltage $v$ and the incident angle $\vartheta_{oil}$.

The raw data acquisition is completed within two separate steps: 1. Recording $N$ polarized TIRF images at a certain TIRF illumination angle under differently linear polarized illuminations; 2. Recording $M$ Ring-TIRF images at different TIRF angles under a cumulatively uniform polarization. In the first step, the scanning galvanometer rotates the incident light, azimuthally varying the position of the incident spot on the retarder while keeping the TIRF angle, i.e., the distance from the incident spot to the retarder's center, unchanged. For instance, as exampled in Fig. 1(b) step1, four polarized TIRF images are recorded at four illumination positions along the dotted red circle. In the second step, the scanning galvanometer completes scanning one circle of the azimuthal angle within one recording's exposure time and sequentially changes the TIRF angle, i.e., the radius of the scanning circle, for each recording. Fig. 1(b) step2 indicates the acquisition of this Ring-TIRF image stack, with varying penetrating depth under varying angles [shown in Fig. 1(c)]. This change of illumination angle change was also referred on the back focal plane [image under one angle is pictured in Fig. 1(d)]. See also a dynamic change of thirteen varying illumination angles in Visualization 1.

Before the experiment, a calibration between the voltage imposed on the scanning galvanometer and the angle from the objective should be done. The relationship is linear in our system. The specific parameters are determined similarly as Ref [29], by sequentially metering a set of angles under certain voltages, as shown in Fig. 1(e). The incident angle in sample is measured as:

$$\theta_{sample} = \arctan(d/h) \quad (2)$$

Thus the angle in the oil immersion objective is:

$$\theta_{oil} = \arcsin\left(n_{sample} \cdot \theta_{sample} / n_{oil}\right) \quad (3)$$

where $n_{sample}$ and $n_{oil}$ refers to the refractive index of the sample and the objective oil respectively. The calibrating results of the measured data is verified in Fig. 1(f) with an R-square of 0.9998. In our experimental setup, the critical angle is 61°. We acquire the polarized TIRF images at 61.5° and the Ring-TIRF image stack at twenty angles starting from 61.5° with an interval of 0.2°.

## B. Forward model

Firstly, we would like to revisit the imaging in conventional TIRF microscopy. Analogous to the experiments, we used a two layer (objective oil and sample) model to describe the evolution of the intensity of fluorescence [36]. When the incident angle of illumination goes beyond the critical angle, total internal reflection occurs at the interface between the substrate and the sample. Simultaneously, an evanescent field is generated, illuminating the fluorescent sample with an exponential decayed axial intensity. At the depth (the distance from the interface of the substrate) of z, the theoretical intensity I is given as:

$$I(z,\theta) = I_0(\theta) \cdot \exp(-z/d(\theta)) \quad (4)$$

where $\theta$ is the incidence angle, $d(\theta)$ is the penetration depth, calculated as:

$$d(\theta) = \frac{\lambda}{4\pi}\left(n_{oil}^2 \sin^2\theta - n_{sample}^2\right)^{-1/2} \quad (5)$$

where $n_{oil}$ and $n_{sample}$ are respective refractive index of the objective oil and the sample. $I_0(\theta)$ is the intensity directly at the interface given by Fresnel's equations and it is polarization related. For p-polarized beam, the value is:

$$I_0(\theta) = \frac{4\cos^2\theta\left(2\sin^2\theta - n^2\right)}{n^2\cos^2\theta + \sin^2\theta - n^2} \quad (6)$$

with $n = n_{sample}/n_{oil}$. Therefore, the obtained TIRF recording under a certain incidence angle $\theta$ comes from the axial intensity integration:

$$y(\theta) = [I_0(\theta)\int_0^\infty S(z)\exp(-z/d(\theta))dz] \otimes PSF \quad (7)$$

where PSF is the detection point spread function of the system, calculated with the vectorial diffraction theory [37] according to our experimental setup. S (z) is the fluorophore emission capacity in the volume.

Then, in polarization modulated TIRF, a polarization modulation term is added when calculating the sample's fluorescence emission. Since the incident light is linearly polarized, the emission intensity is modulated by the relative angle between the incident polarization and fluorophore orientation in the cosine-squared form [35]. As a result, the obtained recording is:

$$y(\theta,\alpha) = [I_0(\theta)\cos^2\alpha\int_0^\infty S(z)\exp(-z/d(\theta))dz] \otimes PSF \quad (8)$$

where $\alpha$ is the relative angle between the incident polarization and the fluorescent molecule orientation. The fluorophore is fully activated only when its dipole orientation is parallel with the incidence polarization. The emission extent periodical variation induced by this polarization modulation term enables the molecules emitting only under certain polarization state, thus imposing sparsity on each polarization-dependent recording.

While in MA-TIRF acquisition, the imaging process in integration format is modified with a linear forward model in a discrete setting. Under N incidence angles, we calculated each image by summing up the intensity at N axial depths, which is indicated by matrix multiplication:

$$y = H \cdot x \quad (9)$$

$$H(\theta,z) = I_0(\theta)\exp(-z/d(\theta)) \quad (10)$$

$$x(z) = S(z) \otimes PSF \quad (11)$$

Here, H is a $N_{angles} \times N_{depths}$ matrix indicating the theoretical incidence intensity in N axial depth under N incidence angles, x and y corresponds to matrixes of size $N_{depths} \times N_{pixels}$ and $N_{angles} \times N_{pixels}$, representing the spatial fluorophore distribution and acquired MA-TIRF image stacks.

## C. Reconstruction algorithm

A flow chart of Pol-TIRF reconstruction process is presented in Fig. 2. Raw image stacks acquired include N TIRF images under N different linear polarizations and M Ring-TIRF images under M illumination angles. The final result in first step is regarded as a spatial support image for the next MA-TIRF reconstruction from the polarization modulated dataset. Due to the low signal-to-noise ratio in TIRF imaging, preprocessing the acquired image stack to extract the general structure of the sample is often adopted for the MA-TIRF reconstruction, i.e., the procedure of our step 2. Here abandoning the conventional diffraction limited spatial support image, a super-resolved structure reconstructed from polarization modulation gives more detail-resolved lateral information. Considering the illumination angle and sample axial distribution is unrelated term in polarization demodulation, we rewrite the imaging model in Eq. 11, including a background noise b:

$$y(\alpha) = \cos^2\alpha \cdot C(\theta,z) \otimes PSF + b \quad (12)$$

$$C(\theta,z) = I_0(\theta)\int_0^\infty S(z)\exp(-z/d(\theta))dz \quad (13)$$

$C(\theta,z)$ is considered as a constant during the reconstruction. After experiencing a Poisson distributed detection procedure, the final acquired fluorescence signal is:

$$\mu(\alpha) \sim Poisson\left[\cos^2\alpha \cdot C \otimes PSF + b\right] \quad (14)$$

The demodulation is accomplished by minimizing a negative Poisson log-likelihood objective function $l(y,\mu) = y - \mu\log y$ to find the optimal solution for C and b. Besides the illustrated sparsity in C, the slowly varying background b also renders a sparse cosine transform. As a result, we added $L_1$ regularization terms according to the character of the two representations to avoid overfitting in this ill-posed inverse problem, leading to the objective function:

$$P(C,b) = \arg\min_{C,b}\{\sum l(y(\alpha);\mu(\alpha)) + \lambda_1\|C\|_1 + \lambda_2\|\tilde{b}\|_1\} \quad (15)$$

Here the objective function is iteratively solved using a fast algorithm FISTA [38], with a convergence speed of $O(1/K^2)$. Although $L_1$

regularization is not differentiable, it can be computational efficiently solved with a soft-thresholding step [39].

Since the three dimensional reconstruction is based on the axial intensity variation of the frames taken under varied angles, the spatial support should not influence the intensity variation tendency. As a result, the laterally super-resolved image is binarized to generate a spatial support for the Ring-TIRF image stack. In the third step, the fixed spatial support is sequentially applied on the Ring-TIRF images, obtaining a matrix $y$ as the input of 3D reconstruction.

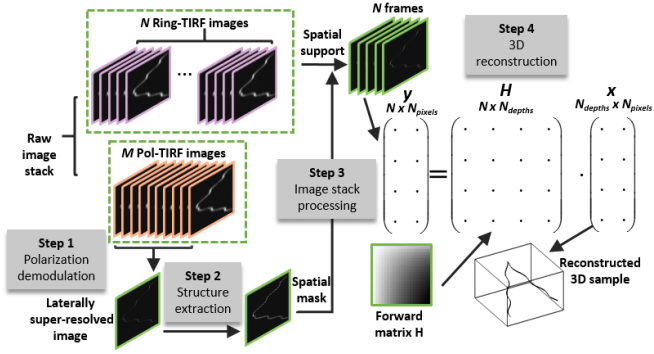

Fig. 2. Flow chart of Pol-TIRF reconstruction. First, the polarization modulated TIRF recordings are demodulated to generate a high resolution image. Then a spatial mask is binarized from the high resolution image. Thirdly, the spatial mask is applied on all the Ring-TIRF images under different illumination angles with the image stack transformed into a matrix of $N_{angles} \times N_{depths}$. This matrix as well as the calculated forward matrix $H$ are the input for the final 3D reconstruction. Finally, the 3D structure of the sample is recovered by iteratively solving the ill-posed inverse problem.

The fourth step is the three dimensional reconstruction problem as expressed by Eq. (8), which can be turned into an optimization problem with the following objective function:

$$\arg\min \|y - H \cdot x\|_2^2 + \lambda_1 \|x\|_1 + \lambda_2 TV(x) \quad (16)$$

The first term penalizes deviations from the forward model and the second and third terms impose regularity on the solution. The second term is an $L_1$ regulation on $x$ considering the positivity of the sample under observation and the third term is an alternative TV regulation, which helps adjusting the result's spatial smoothness. $\lambda_1, \lambda_2 > 0$ are tradeoff parameters that balance the three terms

The main difficulty in the optimization stage stems from the potentially large size of the input data. High accuracy also comes at the cost of a large number of Ring-TIRF angles, as we will show in the following simulation. Our goal then is to handle this large amount of 3D observations at a reasonable computational cost. Here we adopted an algorithm based on the alternating direction method of multipliers (ADMM), which is adapted to the convex and non-smooth form of Eq. (13). Differing from the previous algorithms where the computation load increases quadratically as increasing the amount of input elements, i.e., the input angles, ADMM decompose the original optimization problem into efficiently solvable sub-problems to accelerate the computation [40-42]. Based on variable splitting and Lagrange multipliers, variable $u$ is introduced to replace $x$ and added the last term as a penalty term forcing them to be equal:

$$\arg\min \|y - H \cdot u\|_2^2 + \lambda_1 \|x\|_1 + \lambda_2 TV(x) + \mu\|u - x + \eta\|_2^2 \quad (17)$$

The new optimization problem is solved iteratively by updating one variable at one time. The analytic solution of each variable is obtained explicitly without any iteration, which is the major reason for the fast speed of this algorithm. In iteration $k$, the update steps are:

$$u^k = \left(H^T H + \mu I\right)^{-1}\left[H^T y + \mu\left(x^{k-1} - \eta^{k-1}\right)\right] \quad (18)$$

$$x^k = S_{\lambda_1/\mu}\left[\Psi_{\lambda_2/\mu}\left(u^k + \eta^{k-1}\right)\right] \quad (19)$$

$$\eta^k = \eta^{k-1} + \mu\left(u^k - x^k\right) \quad (20)$$

where $\mu$ is a constant, $S$ denotes the soft-thresholding step and $\Psi$ denotes fixed number of iterations of Chambolle's algorithm for solving TV regulation [43].

To test the efficacy of our method, we conducted simulations of fluorescent samples distributed at different depth. Shown in Fig. 3(a), an array of 25 nm single fluorophores with an axial spacing of 25 nm and a lateral spacing of 160 nm is simulated to test the discerning ability. The acquired image stack are set to be under 50 illumination angles with peak signal to noise ratio (PSNR) of 20 in the raw recording. In Fig. 3(b), the reconstructed fluorophores are clearly separated with an axial full-width-at-half-maximum (FWHM) of 40 nm. Next, the influence of experimental noise and number of illumination angles on the accuracy of the recovery results is evaluated. Fig. 3 (c-d) shows the mean value as well as the range of recovery results under varying imaging conditions, plotted against the real depth of the sample. Fig. 3(c) indicates depth precision improvement is obtained under more illumination angles. In Fig. 3(d), reconstruction accuracy gradually becomes slightly worse when the noise level gets high, especially when the sample locates far from the interface.

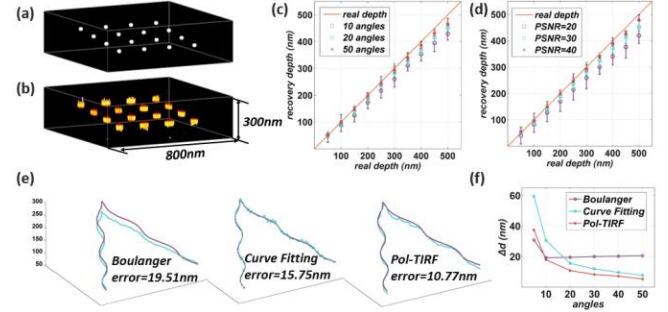

Fig. 3. Simulation results of Pol-TIRF. (a-b) Fluorophore array illustrating discerning ability; (c) Recovery depth plotted against real depth under different illumination angle numbers; (d) Recovery depth plotted against real depth under different PSNR; (e) Ground truth and 3D views of reconstructed results of Boulanger, CF and Pol-TIRF while PSNR=30, illumination angles=20; (f) Standard deviation of difference between the reconstructed results and the microtubule ground truth under different illumination angles.

In addition, we used curvilinear structures generated by trigonometric functions to simulate microtubules and compare our method with the state-of-the-art algorithm in Boulanger's paper [27] as well as the prevailing least squares based Curve Fitting (CF) method [30]. We refer these two methods to Boulanger and CF in the following illustration for simplification. To focus on the axial reconstruction precision, PSF is not considered in the imaging process. Three dimensional views of the ground truth (purple line) and the recovery results (blue line) are presented in Fig. 3(e), where Pol-TIRF exhibits the most correspondence with the ground truth. The Boulanger's result renders a smaller reconstructed depth range than the real value while in CF result, there are several depth jumps making a rough outline though its constructed range is almost the same as the ground truth. The accuracy of the sample recovery, $\Delta d$, is evaluated by calculating the standard deviation of difference between the reconstructed results and the ground truth. When the number illumination angles goes beyond ten, Pol-TIRF shows best results among the three methods, with an axial accuracy around 10 nm.

Theoretically, adopting more input illumination angles would expect a higher axial precision. However, it also means increasing acquisition time. Taking a balance between the acquisition time and algorithm performance, we chose 20 varied illumination angles as a compromise for reconstructing the height image. As a comparison for demonstrating the proposed algorithm's superiority of reconstruction speed, when analyzing a 512x512 image stack with 20 illumination angles, only 3 seconds is required when not applying TV regulation while Boulanger's method took 11 seconds under the same operating environment.

In order to facilitate the study and implementation in TIRF imaging, open-source demo codes for Pol-TIRF can be found in our GitHub repository [44], accompanied with the raw data corresponding to figures in Fig. 4. The code can be used to reproduce the results presented in this paper, or as an accelerated reconstruction algorithm for conventional MA-TIRF as well as a reference for further explorations in TIRF imaging.

## 3. EXPERIMENTAL RESULTS

### A. Resolution illustration

We first imaged a fixed Tubulin sample in Vero cells (Tub-S635P, Abberior) to evaluate the Pol-TIRF resolving ability. Ten polarized TIRF images and twenty Ring-TIRF images were collected under a field-of-view of 32.8 μm ×32.8 μm. As can be seen from the comparisons in Fig. 4(a-b), the filamentous microtubule structures once overlapped in the Ring-TIRF image [Fig. 4(a)] were able to be clearly discerned in polarization demodulated image [Fig. 4(b)]. Also, as we have done in the simulation section, comparison of the reconstruction result with the previously mentioned two method, Boulanger's method and CF, is conducted. The 3D results are turned into 2D colored depth maps by calculating the axial weighted mean value of the 3D result of all the three methods. A conformable microtubule depth tendency is observed in Fig. 4(c-e). The reconstruction shows the microtubules distributed in depth approximately from 20 to 220 nm, where Pol-TIRF has proved to have the highest depth resolution. In addition, same as that indicated in the simulation section, the reconstructed depth range of Boulanger's result is smaller than the other two methods and CF result appears to render a rougher depth variation. Particularly, Pol-TIRF provides the highest transverse resolution, which is clearly seen from Fig. 4(h), the amplified views indicated by the red box areas in Fig. 4(c-e). A transverse line profile across the red dashed line shows a transverse resolution of 180 nm [Fig. 4(g)]. Finally, an axial resolution of 40 nm [Fig. 4(f)] for the microtubule located at depth around 180 nm in the red box proved to be higher than that of 50 nm evidenced in Boulanger method, evidencing the superiority of the proposed method.

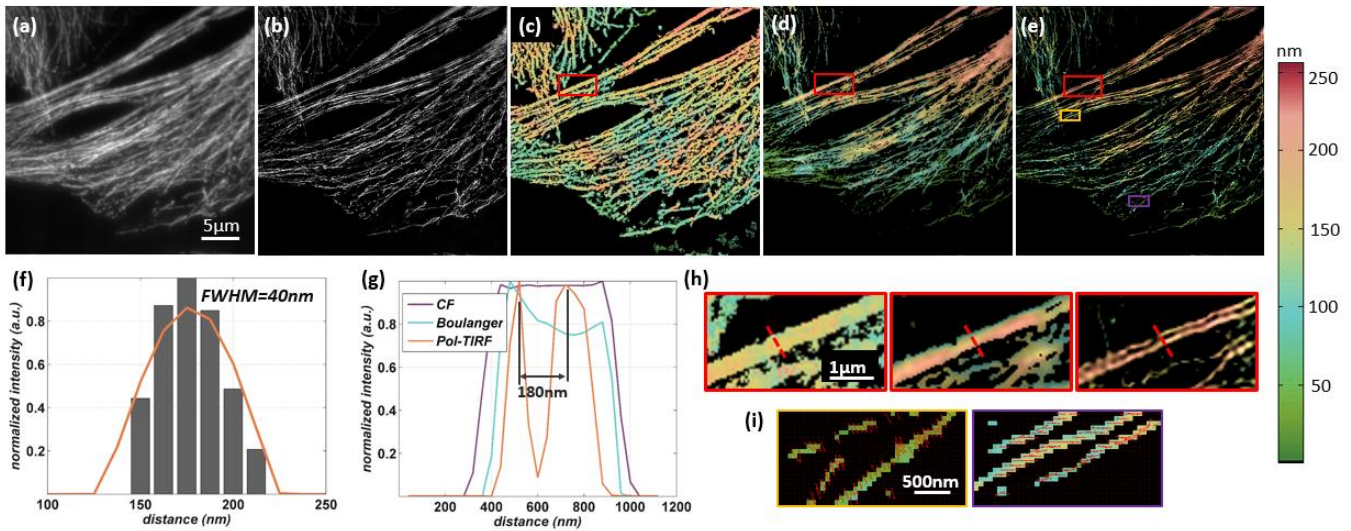

Fig. 4. Pol-TIRF imaging of fixed microtubules. (a) Single Ring-TIRF image; (b) Polarization demodulated image; (c-e) Depth map reconstructed with Curve Fitting, Boulanger's method and Pol-TIRF; (f)Axial distribution of microtubule located at around 180 nm depth showing an axial resolution of 40 nm; (g) Line profile across the dashed red line in (h) illustrating a transverse resolution of 180 nm; (h) Magnified views of regions marked by red box in (c-e); (i) Fluorophore dipole orientation map of selected areas in (e).

Moreover, as the bonus of polarization modulation, Pol-TIRF provides additional information of fluorophore dipole's orientation that other method can never achieve [33]. Using the method detailed in Ref 33, we superimposed the orientation map onto the depth map, in which the red arrows denote the effective dipole orientation and the length denotes the confidence of the result. Particularly, as can be seen in Fig. 4(i), we found that on sparsely distributed samples, such as the microtubules here, the dipole orientations are vertical to the microtubule direction while they turn parallel to the microtubule direction in densely distributed samples.

We next imaged fixed nuclear pore complex in Vero cells (Nup153-STAR635P, Abberior Instruments) with a total volume of 32.8 μm × 32.8 μm × 250 nm, as shown in Fig. 5, where the areas for y-z view and x-z view are respectively indicated by the dashed purple box and orange box in Fig. 5(a), which clearly exhibits the axial distribution. Interestingly, we found the center part of the cell is very flat while there was a collapse in its both ends, with thickness of the whole cell no larger than 150 nm.

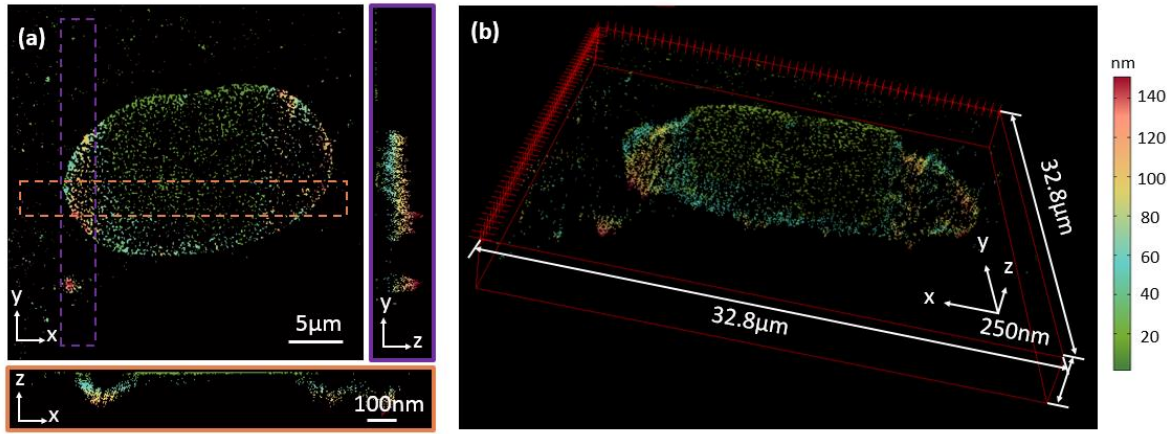

Fig. 5. Pol-TIRF imaging of fixed nuclear pore complex. (a) Pol-TIRF reconstructed depth map in x-y view, x-z view and y-z view. Scale bars in x-y view and x-z view, represent the unequal scales in x- y axis and z axis, respectively; (b) 3D view of the reconstructed cell distribution.

## B. Cell dynamics

Owing to the high time resolution, Pol-TIRF is capable of illustrating the three dimensional dynamic changes of live cells. Mitochondrion, with a size between 0.75 and 3 μm in diameter, is a dynamic organelle that participate extensively in many important cellular processes. ATTO 647N (ATTO-TEC) has been recently reported to have high fluorescence intensity and excellent photo stability as mitochondrial marker in live cell imaging [45]. Resorting this dye, we observed mitochondria in live U20S cells for detailed structures. Here, setting the exposure time for each image is 50ms and taking the response time of electronic devices into account, the acquisition time of one image stack is approximately 2s.

Several granular mitochondria, rod-like mitochondria and mitochondria with other shapes were observed in Fig. 6. The conventional Ring-TIRF image in Fig. 6(a) is blurred with strong fluorescent background while in Pol-TIRF image [Fig. 6(b)], removal of the background and resolution enhancement enables the vision of mitochondria cristae. These can be more clearly examined in the magnified views of the sub areas in the white box. Axial and transverse profiles of a typical tubular structure in mitochondria, marked with a red line, are plotted in Fig. 6(b), showing an axial resolution of 80 nm and a transverse resolution of 140 nm, respectively. The result here resonates well with the previous measuring of the inter mitochondria tubules width using stochastic optical reconstruction microscopy (STORM) [46]. Further, we took a sequence of 20 image stacks with a time interval of 2s, where consecutive processes of active mitochondrial fusion and fission were successfully rendered (Visualization 2). As evidenced, the time-lapse Pol-TIRF depth maps of the selected subarea [Fig. 6(c)] revealed thin, extended tubular intermediates connecting neighboring mitochondria both prior to fission and after fusion. For instance, as vividly shown in this 40s time-lapse, the organelle marked by the purple triangle split into two mitochondria while the two marked by the red triangle merged into one conversely, with a clear height map further characterizing the movement of the process (see also about the video dynamics in Visualization 2).

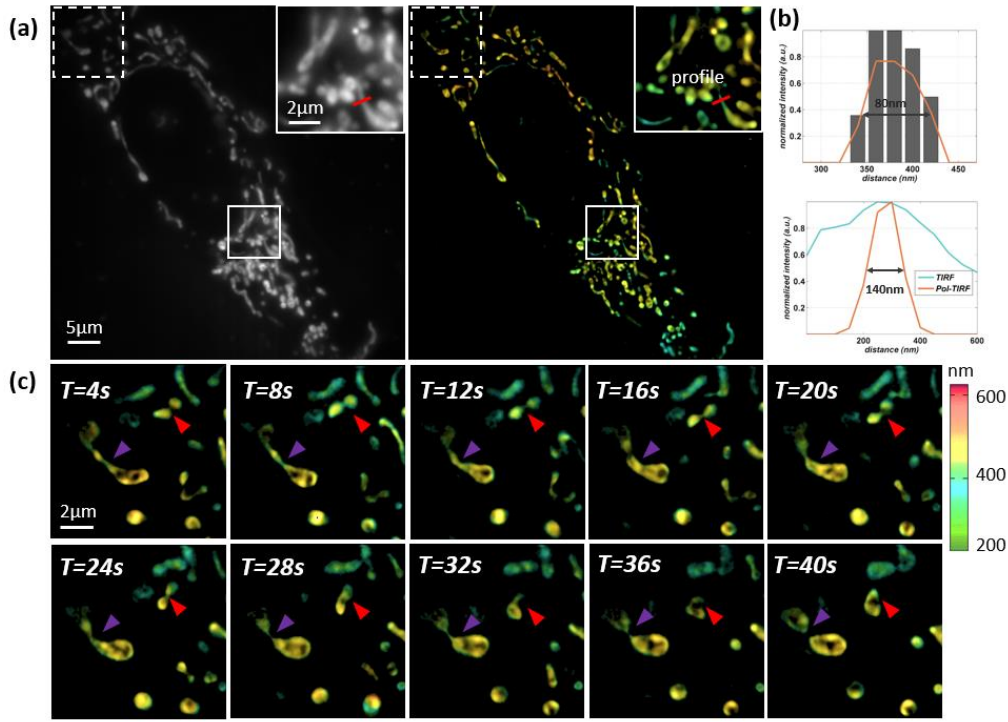

Fig. 6. Mitochondria dynamics in a live U20S cell. (a) Single Ring-TIRF image and Pol-TIRF reconstructed colored depth map; (b) Axial and transverse profile along the red line in (a), respectively; (c) Fission (purple triangle) and fusion (red triangle) events in the dotted white box in (a) captured by a time-series of 2-sec Pol-TIRF image stacks.

## 4. CONCLUSION

In summary, we have developed a simple strategy based on Pol-TIRF to determine the three dimensional structure of cell proteins with isotropically enhanced resolution in living cell. Given the results above, it would be reasonable to conclude the Pol-TIRF outperforms the existing algorithms and setups upon MA-TIRF imaging. Being a very first implementation, the polarization modulation exampled here proves to be a simple way of providing extra lateral information for the MA-TIRF imaging. Surely, beneath the concept explored here, one could also find another potential scheme to offer an enrichment in transverse information, e.g., TIRF-SIM in which an additional pair of laser beams controlled by scanning galvanometer or a spatial light modulator should be introduced to interfere out a structured illumination in the focal plane[47, 48]. Having a potential to improve the lateral resolution of two folds, this potential scheme in turn largely increase the system complexity and alignment difficulty, when combining with MA-TIRF. While in our instance, the only additional device is a special vortex half-wave retarder to alter the polarization state of the illumination beam, which is rather compatible with the current scanning TIRF illumination path. Hence Pol-TIRF would be the probably the most efficient way to achieve the goal of promoting the resolution while keeping the tight system, and thus we are faithful it would be quickly adopted by the TIRF imaging family.

In the development of our algorithm, we also leveraged state-of-the-art techniques in signal processing and optimization including FISTA and ADMM, which separately enhance the convergence and reconstruction speed. The robustness of the proposed method has been extensively studied by simulation and experimental data in comparison with two previous algorithms. We hope that the Pol-TIRF here can provide a different prospective in developing both the TIRF algorithms and the imaging system since it proves advantages in these two aspects. Moreover, the polarization modulation highlighted here could also inspire other techniques, such as the single molecule localization methods, where the effect of illumination polarization has already been explored [49, 50], but not for directly resolution promoting purpose. Our future work in this area will investigate further adaptations including promoting the discrimination of dipoles in single molecule localization methods and so on.

**Funding**. National Basic Research Program of China (973 Program) (2015CB352003); National Natural Science Foundation of China (61735017, 61427818, and 61335003); The National Key Research and Development Program of China (2017YFC0110303, 2016YFF0101400); NSFC of Zhejiang province (LR16F050001); Fundamental Research Funds for the Central Universities (2017FZA5004).

**Acknowledgment**. We thank Dr. Meng Zhang and Prof. Yu-Hui Zhang from Huazhong University of Science and Technology, China for their help with the live U2OS cell sample. We also thank the useful discussions from Dr. Yujia Huang from Caltech, America.